\documentclass[a4paper,11pt]{article}
\pdfoutput=1 

\usepackage{jheppub} 

\usepackage[T1]{fontenc} 

\newcommand{\avg}[1]{\left< #1 \right>} 

\title{Asymptotic critical behavior of holographic superconductor phase transition - 
the spectrum of excited states becomes continuous at $T=0$}

\author[a,1]{Toan T. Nguyen,\note{Corresponding author.}}
\author[b]{Tran Huu Phat}

\affiliation[a]{Key Laboratory for Multiscale Simulation of Complex Systems 
and Department of Theoretical Physics,
University of Science, Vietnam National University -- Hanoi, 
334 Nguyen Trai street, Thanh Xuan, Hanoi 100000, Vietnam
}
\affiliation[b]{Vietnam Atomic Energy Commission, 
59 Ly Thuong Kiet street, Hoan Kiem, Hanoi 100000, Vietnam
}

\emailAdd{toannt@hus.edu.vn}
\emailAdd{thphat@live.com}

\abstract{
Within the framework of AdS/CFT duality, excited states of the conformal field living 
at the global AdS boundary of a four-dimensional spacetime Einstein gravity are investigated analytically
in the probe limit where the field equations are linearized.
At asymptotically large values, the threshold chemical potential for the appearance of excited condensate states are discrete, equal spacing, with the gap approaches zero logarithmically in the limit $T\rightarrow 0$.
Remarkably, numerical results show that, this behavior applies even for states
as low as for the first or the second excited state of the condensate.
This is especially significant on the liquid side of the black hole van der Waals - like phase transition (small or zero topological charge) where there seems to be no gap between the ground state and the first excited state
at zero temperature. 
We postulate that, at the exact limit $T = 0$ where the gap is zero,
the spectrum of threshold chemical potentials becomes continuous,
all excited states of the condensate are activated above a finite chemical potential, 
suggesting a new quantum phase transition as a function of the chemical potential. 
Previous studies have largely missed this continuous spectrum of excited states in the $T\rightarrow 0$ limit.
This fact should be taken into account carefully in AdS/CFT duality studies.
}

\begin{document} 
\maketitle
\flushbottom

\section{Introduction}

The AdS/CFT duality \cite{maldacena1998large,witten1998anti,gubser1998gauge}
has opened up a new direction of
connecting gravity with all other branches of physics,
offering a powerful formalism for studying them. 
In this respect, the study of holographic phase transitions has
been developed and gained great successes associated with superconductors
\cite{hartnoll2008holographic,basu2009supercurrent,herzog2009holographic}. 

Over the past few years, there has been a growing interest in studying excited states of the holographic condensate living at the global AdS boundary of a bulk Einstein gravity \cite{gubser2008colorful,wang2020excited,bao2021excited,pan2021holographic}.
These excited state solutions have multiple nodes (zeroes) and higher energies than the ground state.
Although, their role in the black hole system is still unclear, 
they are of great interest for superconductor studies in particular, and in condensed matter physics in general.
Through the AdS/CFT duality, one is able to understand behaviors of excited states of
many types of superconductors using different models of black holes.

Most studies of excited states in AdS/CFT duality so far have relied on numerical approaches,
and have troubles extending to zero temperature due to the non-analyticity of the field equations. 
In this work, the limiting behavior of these excited states is solved analytically in the probe limit where
the field equations can be linearized.
It is shown that the spectrum of excited states becomes continuous at zero temperature, 
a fact which has been largely missed in previous studies. 
Our system of investigation is a model of 4D Einstein-Maxwell blackhole at finite topological charge \cite{Tian_2019}, 
but the results can be easily expanded to other systems and other dimensions.

We start from the model of an Abelian Higgs field and a Maxwell field in the four-dimensional spacetime Einstein gravity. The bulk action is given by,
\begin{gather}
S = \frac{1}{16\pi G} \int d^4x \sqrt{-g} \left( R - \frac{16}{L^2}
    -\frac{1}{4}F_{\mu\nu}F^{\mu\nu} 
     - | \nabla \Psi - i q A \Psi |^2 - m^2 |\Psi|^2
    \right) ~,
\label{eq:S}
\end{gather}
where $G$ is the Newton constant. In the uncondensed phase, the solution to Eq. (\ref{eq:S}) are the Reissner–Nordstr\"{o}m black hole (BH):
\begin{equation}
ds^2 = -f(r) dt^2 + \frac{dr^2}{f(r)} + r^2 d\Omega_{2,k} ,
\end{equation}
where
\begin{equation}
 f(r) = k - \frac{2M}{r} + \frac{Q^2}{r^2} + \frac{r^2}{L^2}    ,
\end{equation}
and $d\Omega_{2,k}$ is the metric of the two-sphere of radius
$1/\sqrt{k}$. 
The BH event horizon $r_0$ is the largest solution of the equation
$f(r_0)=0$. The BH Hawking temperature $T$ is:
\begin{eqnarray}
T_{\mbox{\scriptsize Hawking}} &=& \frac{f'(r_0)}{4\pi} ,
\end{eqnarray}
which is also regarded as the temperature of the conformal field at the AdS boundary.

It is well-known that there are some deep connections between black holes and thermodynamics.
One frequently uses the concepts of black hole entropy, temperature and pressure
and obtains many interesting correspondence between the two objects.
While the flat horizon geometry with $k=0$ is more common in AdS/CFT duality studies,
the inclusion of non-flat horizon geometries where $k \neq 0$ 
offers a complete phase space where the mass $M$ of the blackhole is no longer regarded
as the internal energy, but rather the enthalpy \cite{kastor2009enthalpy}
(A review of this expanded thermodynamics can be found in Ref. \cite{kubizvnak2017black}).
Specifically, for $k>0$, using the entropy $S=\pi r_0^2$, and the pressure $P=3/8\pi L^2$,
there is a total analogy between the small$-$large BH phase transition 
and the liquid–gas phase transition of the van der Waals (vdW) theory \cite{ammon2015gauge}. 
In this set up, $M$ is the enthalpy and $k$ is interpreted as the measure of a new charge, 
the topological charge. 
For $k=1$, $L<6$ corresponds to the liquid side, while $L>6$ corresponds to the gas side of the vdW transition
\cite{phat2021holographic}.
For this reason, even though this work focuses on understanding 
asymptotic behavior of the comformal field at large chemical potential,
the case of finite topological charge is explicitly studied in the model.
This allows its expanded application to a more complete thermodynamic phase diagram of black holes.

Using the ansatz, $A_\mu = (\Phi(r), 0, 0, 0)$, $\Psi = \Psi(r)$,
minimizing the action with respect to $\Phi(r)$ and $\Psi(r)$,
one arrives at the equations of motion for the fields:
\begin{gather}
\Psi''+ \left(\frac{2}{r} + \frac{f'}{f} \right) \Psi' 
        + \left(\frac{q^2\Phi^2}{f^2}+\frac{2}{L^2f}\right) \Psi = 0 ,
        \\
\Phi''+ \frac{2}{r} \Phi' 
        - \frac{2q^2|\Psi|^2}{f} \Phi = 0 .
\end{gather}
At the AdS boundary $r\rightarrow \infty$,
%
\begin{eqnarray}
\Phi(r \rightarrow \infty) &=& 
    \mu - \frac{\rho}{r} + .... \\
\Psi(r \rightarrow \infty) &=& 
    \frac{\Psi^{(1)}}{r} + \frac{\Psi^{(2)}}{r^2} + ...
\end{eqnarray}
At large temperature, the system has only a ``trivial'' normal solution:
\begin{eqnarray}
\Phi(r) &=& \mu \left( 1 - \frac{r_0}{r} \right) , \\
\Psi(r) &=& 0 .
\end{eqnarray}
As the temperature decreases below a critical value $T_c$,
the system permits a non-vanishing $\Psi(r)$ condensate solution. 
For the dual holographic superconductor at the AdS boundary, 
the condensate solution with $\Psi^{(1)} \neq 0$, $\Psi^{(2)}=0$ is called $\avg{{\cal O}_1}$ condensate.
Similarly, the condensate solution with $\Psi^{(2)} \neq 0$, $\Psi^{(1)}=0$ is called $\avg{{\cal O}_2}$ condensate.
In the dictionary of AdS/CFT duality,
$\mu$ is intepreted as the chemical potential of the conformal field living at the boundary, $\rho$ is the corresponding density \cite{herzog2009holographic}.
In this work, we will work in the grand canonical ensemble where the chemical potential $\mu$ is fixed.

Depending on the value of the chemical potential, as the temperature
decreases even further, other solutions become permissible, which
are identified as the 1st excited state, 2nd excited state, ... 
of the condensate \cite{wang2020excited}. 
This work attempts to understand critical properties of these states, 
especially the asymptotic behavior 
at large chemical potential $\mu$.

Without losing generality, we set $Q=q=1$.
To simplify formulas, the coordinate is changed to $z = r_0/r$, and dimensionless quantities are used:
\begin{gather}
f(z) = \frac{f(r)}{z_h^2} = \frac{k}{z_h^2} 
    -  \frac{2 M}{z_h} z + z^2 + \frac{1/L^2z_h^4}{z^2} , \\
\Phi(z) = \Phi(r=1/z)/z_h^3 , \\
\Psi(z) = \Psi(r=1/z) / z z_h^2 ,
\end{gather}
here
\begin{gather}
z_h=\frac{1}{r_0}, \quad 
T = \frac{4\pi  T_{\mbox{\scriptsize Hawking}}}{z_h^3} = - f'(1) .
\end{gather}
$\Psi^{(1)}$, $\Psi^{(2)}$, $\mu$, $\rho$ are also rescaled accordingly (not listed here to avoid crowding). 
In the rest of the paper, the dimensionless version of these quantities is always assumed unless explicitly stated otherwise.
The AdS boundary is now at $z=0$. The BH event horizon is at $z=1$
which is the smallest solution of $f(z)=0$. 

Notice that, for convenience, the field $\Psi(r)$ is not only rescaled, but also changed. 
With the new definition, the boundary conditions at the AdS boundary are:
\begin{eqnarray}
\Psi(z \rightarrow 0) &=& \Psi^{(1)} + \Psi^{(2)} z + \cdots , \\
\Phi(z \rightarrow 0) &=& \mu - \rho z + \cdots .
\end{eqnarray}
Because $z=1$ is the smallest positive solution of $f(z)$, 
the latter can be rewritten in a more intuitive form:
\begin{equation} 
f(z) = \frac{z-1}{z^2} \left[ (z-1) (z^2 + 2bz + b) - z^2 T \right ] ,
\label{eq:frcompact}
\end{equation}
where $b=1/L^2z_h^4$. 
The dimensionless equations of motion are:
\begin{gather}
\Psi'' + \frac{F^{'}}{F} \Psi' 
    + \frac{zF ^{'} - 2F + 2b}{z^2F} \Psi 
    + \frac{\Phi^2}{F^2} \Psi = 0 ,
\\
\Phi'' - 2 \frac{\Psi^2 \Phi}{F} = 0 ~,
\end{gather}
with $F(z) = z^2 f(z)$.
The ``trivial'' normal solution of the equation of motion now reads
\begin{eqnarray}
\Phi(z) &=& \mu (1-z) , \\
\Psi(z) &=& 0 .
\end{eqnarray}
\begin{figure}[htp]
\centering
\includegraphics[width=0.95\textwidth]{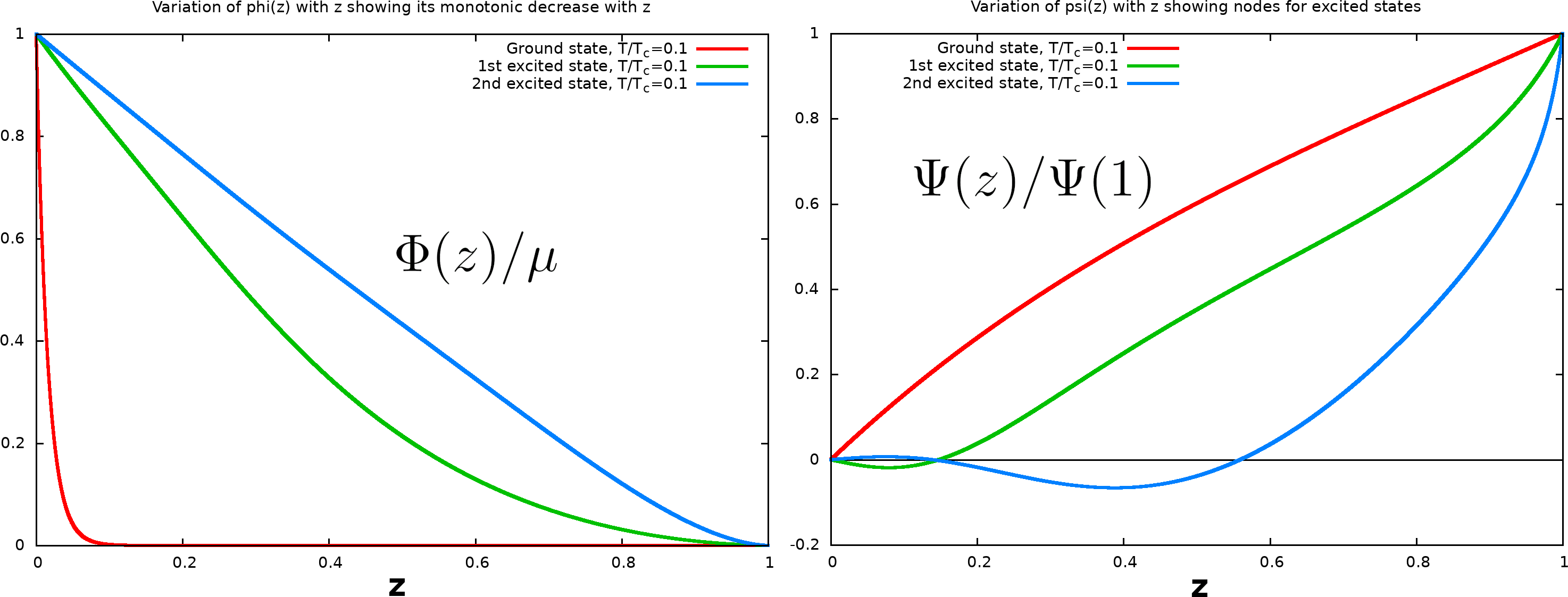}
\caption{Numerical solutions of equations of motion for $\avg{{\cal O}_1}$ condensate at $\mu=8$, $L=1$, $k=1$, and at temperature $T/T_c = 0.1$ for the field $\Phi(z)$ (left) and 
$\Psi(z)$ (right) as functions of $z$.
There are three different condensate solutions with this set of parameters: ground state solution (red),
first excited state solution (green), second excited state solution (blue).
}
\label{fig:Tc}
\end{figure}
For condensate solution at low temperature, the expectation value of the operator, $\avg{{\cal O}_\delta} =\sqrt{2} \Psi^{(\delta)}$ ($\delta = 1$ or 2) is non-zero. 
%
Analytical expressions are not available even for simplest forms of the metric, $f(z)$, 
and one typically resorts to numerical solution.
In Fig. \ref{fig:Tc}, condensate solutions for $k=1$, $L=1$,
$\mu=8$, and $T=0.1 T_c$ are shown as a typical example. 
As the temperature decreases
below the critical temperature $T_c = 1.699$, the condensate ground state is available. 
Below the critical temperature $T_c = 0.295$, the first excited state is available, the field $\Psi(z)$ has one node in the interval $(0,1)$. 
Finally, below the critical temperature $T_c = 0.0806$, the second excited state is available,
$\Psi(z)$ has two nodes in the interval $(0,1)$.

Nevertheless, in numerical solutions, it is easy to overlook various hidden symmetries in the system. 
In this paper, we attempt to investigate behavior of the system near the critical temperature (the probe limit) 
using semi-analytical methods, which reveals several interesting features of the system.

The present paper is organized as follows. 
Section \ref{sec:Tc} deals with the critical transition temperature
of the holographic superconductor phase transition, and the corresponding critical field equation.
Section \ref{sec:muc} is devoted
to a combined analytical and numerical investigation of
the threshold chemical potential for the appearance of various states of the condensate,
especially proof of the continuous spectrum of excited states at $T= 0$.
Section \ref{sec:disc} investigates various aspects of the theory, how it can be expanded to
other types of BH, other dimension, ... .
Section \ref{sec:criticalExp} deals with 
the critical exponent of the order parameter of the holographic
phase transition in the ground as well as excited states.
Finally, the conclusion and outlooks are given in the last section,
Sec. \ref{sec:conclusion}. 

\section{The field equation at criticality and the critical temperature $T_c$}
\label{sec:Tc}

First, let us see how the system undergoes phase transition toward condensate state as
the temperature is lowered. Specifically, 
the first question one asks is, given the chemical potential $\mu$, how many condensate solutions are permitted, and 
what are the corresponding critical transition temperatures?
Near critical temperature, $\avg{{\cal O}_\delta}$ are small. 
one does not need to solve the full set of equations of motion
to obtain the critical temperature \cite{siopsis2010analytic}. 
This analysis also 
provides a check on our formulation by comparing with previous results obtained by directly solving the full system of equations of motion.

Indeed, in the probe limit (or near critical transition temperature),  $T\approx T_c$,
the condensate solution is very small, $\Psi(z) \approx 0$, and $\Phi (z) \approx \mu (1 - z)$. 
The equation for $\Psi$ becomes linearized:
\begin{gather}
\Psi'' + \frac{F^{'}}{F} \Psi'  
    + \frac{ zF ^{'} - 2F + 2b }{ z^2 F} \Psi
    + \frac{ \mu^2(1-z)^2 }{ F^2 } \Psi= 0 .
\label{eq:muTc}
\end{gather}
%

Equation (\ref{eq:muTc}) is to be solved with specific boundary conditions depending on the type of the condensate. 
At a given $\mu$, the procedure for numerical calculation of the critical transition temperature $T_c$ is following.
Since the equation is linear, 
one sets $\Psi(1)=1$, $\Psi'(1) = (2b-T)/T$. 
Numerically integrating the equation from $z=1$ to $z=0$, 
one obtains unique boundary values $\Psi(0) = \Psi^{(1)}(\mu,T)$ 
and $\Psi'(0) = \Psi^{(2)}(\mu,T)$ as functions of given $\mu$ and $T$.
For $\avg{{\cal O}_1}$ condensate, for example, 
by varying $T$ at fixed $\mu$, every time $\Psi^{(2)}(\mu,T)$ crosses zero, 
the corresponding $T$ is the critical temperature of a condensate state.

%
\begin{figure}[htp]
\centering
\includegraphics[width=0.9\textwidth]{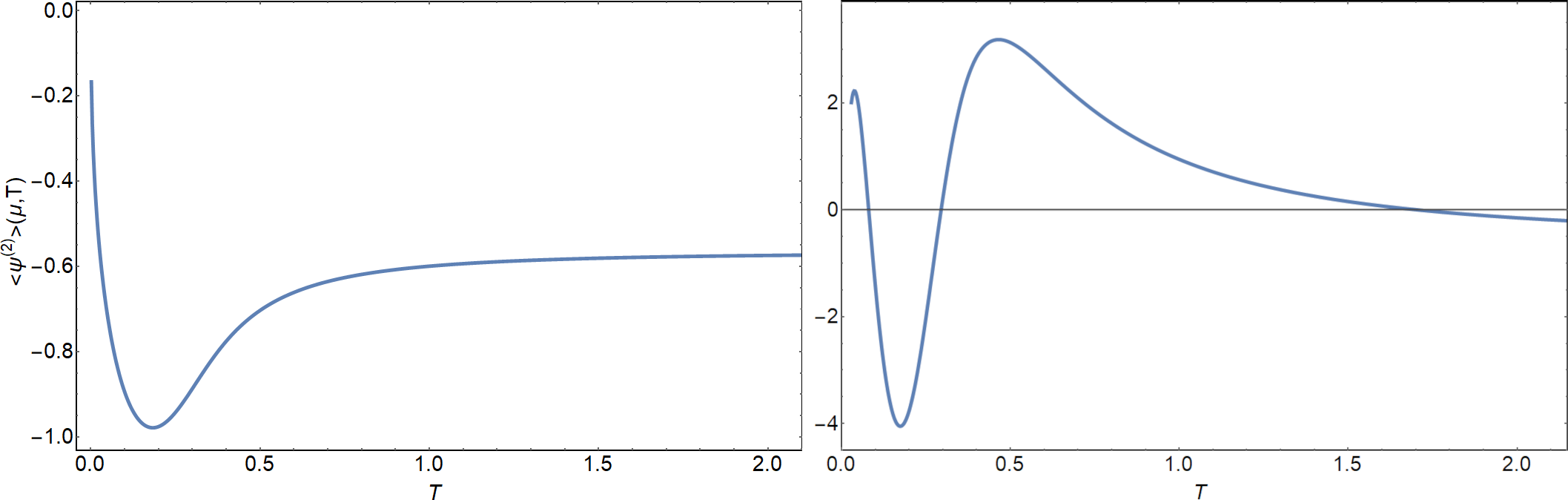}
\caption{
$\Psi'(0)=\Psi^{(2)}(\mu,T)$ as functions of $T$ from numerically solving Eq. (\ref{eq:muTc}) for $\mu = 0.3$ (left) and $8$ (right) for $k=1$, $L=1$.}
\label{fig:O1vsT}
\end{figure}
As an example, Fig. \ref{fig:O1vsT} shows $\Psi^{(2)}(\mu,T)$ as function of $T$ for the case $k=1$, $L=1$, for two different values of the chemical potential $\mu =$ 0.3 and 8.
At $\mu=0.3$, $\Psi^{(2)}(\mu,T)$ never crosses zero. This means that the system is always in the normal state, no phase transition at any finite temperature $T$. 
At $\mu=8$, $\Psi^{(2)}(\mu,T)$ crosses zero three times.
This means that, at this chemical potential, the system goes through phase transition to the ground, 1st excited, and 2nd excite states 
at $T_c =$1.699, 0.295, and 0.0806 respectively. 
These values exactly match critical temperatures obtained by directly solving the full system of equations of motion \cite{phat2021triplet} (see also Fig. \ref{fig:Tc}).
Here, they are obtained through a much simpler method, by solving only one linear equation.

\section{\label{sec:muc}The threshold chemical potential and its asymptotic behavior}

As demonstrated, for small $\mu$, the system permits no condensate solutions. As $\mu$ increases beyond a threshold $\mu_0$, a condensate solution is permitted below a certain temperature $T_c$ where the expectation value of the operator is non-zero, 
$\avg{{\cal O}_\delta} \neq 0$. 
This $T_c$ is the critical temperature of the holographic transition.
As $\mu$ increases further, more solutions are permitted at lower $T_c$ which one identifies as excited states of the condensate. 
In the $n^{\mbox{th}}$ excited state, the field $\Psi(z)$ has $n$ nodes in the interval $z \in (0,1)$.
Therefore, the next question one would like to ask is, at which threshold value $\mu_0$, the ground state solution is available; at which threshold value $\mu_1$, the first excited state solution is available, etc.

Clearly, at each threshold $\mu$ where a new solution (new condensate state) appears, its critical temperature $T_c$ is zero. As $\mu$ increases, $T_c$ increases from zero. Therefore, we seek
to investigate the criticality $\Psi(z)$ equation, Eq. (\ref{eq:muTc}), in the limit $T\rightarrow 0$.
Before going to analytical discussions, let us give 
typical examples of threshold $\mu$ behaviors 
obtained from numerical solution.
\begin{figure}[ht]
\centering
\includegraphics[width=0.9\textwidth]{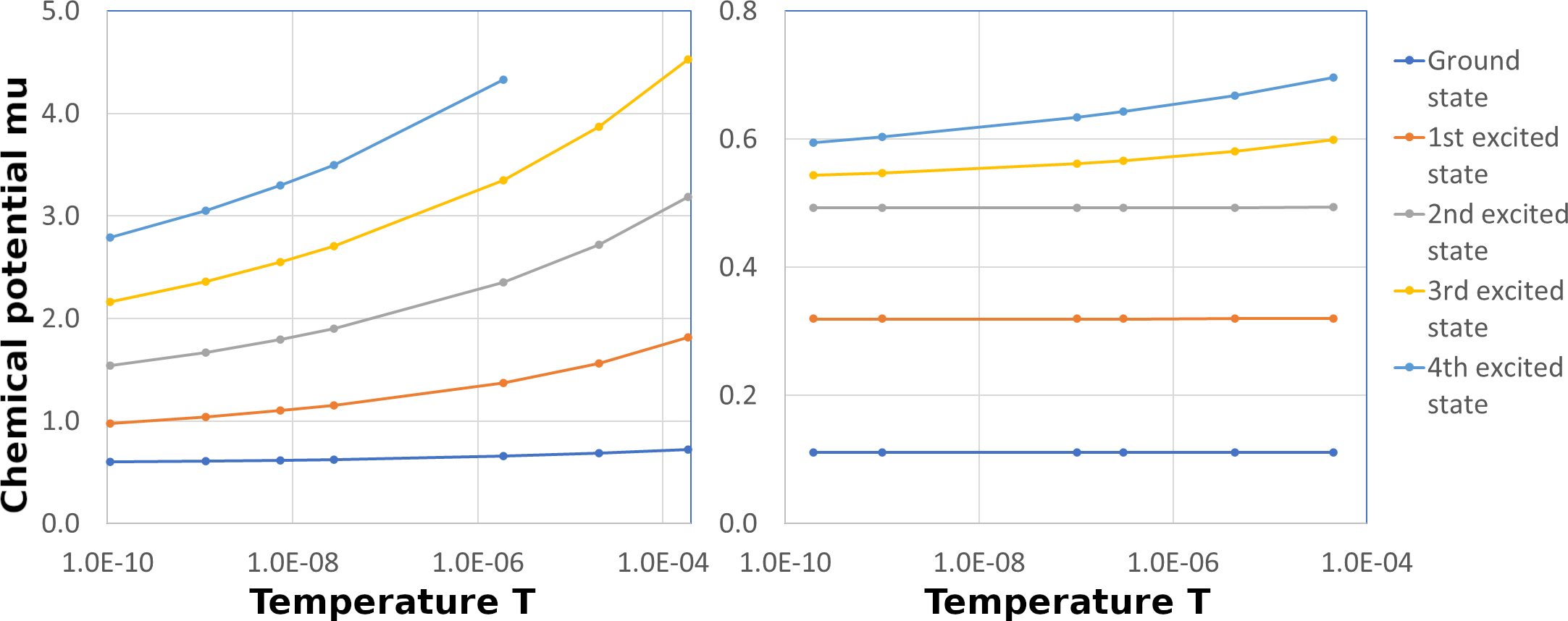}
\caption{
The chemical potential $\mu$ where the ground and first four excited states appears when $T$ decreases, plotted for two cases $L=1$ (left) and $L=9$ (right). 
The threshold $\mu$ are limits of these curves at $T\rightarrow 0$.
Notice the horizontal log scale, and how $\mu$ does not plateau yet for many excited states, 
especially in the $L=1$ case. 
}
\label{fig:mucvsT}
\end{figure}
Fig. \ref{fig:mucvsT} shows the threshold $\mu$ for five lowest energy states as the temperature $T$ decreases, for $L=1$ and $L=9$.  In both cases, the topological charge $k=1$. 
One sees that the threshold $\mu$ for ground state, and maybe first few excited states, shows saturation.
However, for higher excited states, $\mu$ shows no signs of saturation for $T$ as low as $10^{-10}$.

To understand better this behavior, we resort to analytical discussions. Rewrite Eq. (\ref{eq:muTc}) as:
\begin{equation}
- (F \Psi')'  - \frac{zF ^{'} - 2F + 2b}{z^2} \Psi 
    = \mu^2 \frac{(1-z)^2}{F}\Psi .
\label{eq:SLstandard}
\end{equation}
This is the Sturm$-$Liouville (S$-$L) eigenvalue problem. 
Its spectrum consists of an infinite number of eigenvalues which are real, countable, bounded from below and unbounded from above (more on this property later):
\begin{equation}
\mu_0^2 < \mu_1^2 < \mu_2^2 < \cdots \quad \mbox{ with } 
    \lim_{n\rightarrow \infty}  \mu_n^2 \rightarrow \infty ~ .
\label{eq:munlimit}
\end{equation}
%
The corresponding eigenfunction, $\Psi_n(z)$, has exactly $n$ zeros in the interval $(0,1)$. 
Therefore, one easily identifies $\mu_n$ as the threshold chemical potential 
above which the system permits at least $n$ different solutions: 
the ground state and $n-1$ excited states.
For $\mu < \mu_0$, no non-trivial solutions are available,
the system cannot undergo phase transition and
the normal state prevails all the way to the $T=0$ limit.
The rest of this paper concerns with explicit estimates for these eigenvalues. 

We introduce new coordinate $x$ and field $\omega(x)$:
\begin{gather}
x = \int_0^z~ds~ \frac{1-s}{F(s)}, 
\label{eq:TvsZ}
\\
\omega (x) = \Psi(z) \sqrt{1-z} ,
\end{gather}
%
the S$-$L equation becomes,
\begin{gather}
- \omega''(x)
+ V(x) \omega(x) = \mu^2 \omega(x) ,
\label{eq:SLSchro}
\end{gather}
which is exactly the one-dimensional Schr\"{o}dinger equation for an unit mass particle 
moving in an ``effective potential'':
\begin{gather}
V(z(x))	= - \frac{ z [ 2 (z -1) +2b - T ]F}{(1-z)^2} 
		 - \frac{1}{2} \frac{1}{(1-z)^2} \left\{ \frac{3F^2}{2(1-z)^2} 
		 + \frac{F_z F}{1-z}  \right \} ,
\end{gather}
here $F_z$ is the derivative with respect to the variable $z$. 

The Schr\"{o}dinger equation is to be solved with appropriate boundary conditions.
Because $F(z) > 0$, the relation, Eq. (\ref{eq:TvsZ}), gives one-to-one mapping between $x$ and $z$.
The interval $z \in [0,1)$ maps into $x \in [0,c)$ where
\begin{equation}
c = \int_0^1~ds~ \frac{1-s}{F(s)} ,
\end{equation}
is positive and finite at finite $T$, and grows logarithmically with $1/T$ as $T\rightarrow 0$. 
For analytical solution, it is more appropriate to specify boundary conditions
at the regular $x=0$ boundary, requiring only that $\Psi(1)$ is finite 
(remind that we are working with a linear equation).
Thus, $\omega(c) = 0$, a common physical requirement
for Schr\"{o}dinger equations at singular boundaries.

Boundary conditions at $x=0$ depend on the type of condensate operator one wishes to study:
either $\Psi(0)=1$ and $\Psi'(0)=0$ for $\avg{{\cal O}_1}$ condensate; or $\Psi(0)=0$ and $\Psi'(0)=1$ for $\avg{{\cal O}_2}$ condensate.
These transform into corresponding boundary conditions for the $\omega(x)$ field:
\begin{gather}
\omega(0) = \Psi(0) = 1 \mbox { or } 0 , \\
\omega'(0) = b \left[ \Psi'(0) - \frac{1}{2} \Psi(0) \right]  
    = - \frac{b}{2} \mbox { or } b .
\end{gather}
for $\avg{{\cal O}_1}$ or $\avg{{\cal O}_2}$ condensate.

With the familiar Schr\"{o}dinger equation, Eq. (\ref{eq:SLSchro}), 
one understands much better physical behaviors of the system using known results 
from non-relativistic quantum mechanics. %
\begin{figure}[ht]
\centering
\includegraphics[width=0.9\textwidth]{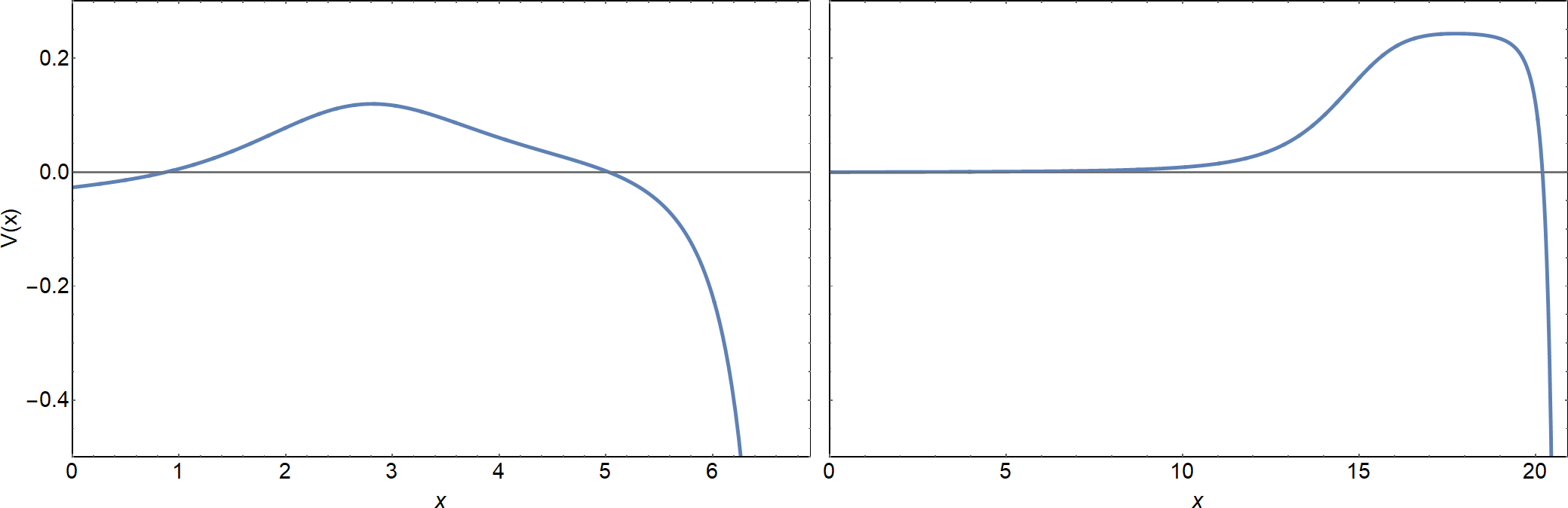}
\caption{The typical behavior of the ``effective potential'' $V(x)$ at low temperature for $L=1$ (left) and $L=9$ (right). 
At $x\rightarrow 0$,
$V(x) \simeq -\frac{3}{4}b^2+b^2(1-2b)x$, which is negative. 
At $x\rightarrow c$, it is again negative and diverges as $-(c-x)^{-2}/4$. 
Because $V(x)$ is continuous, it must be bounded from above.
}
\label{fig:qt}
\end{figure}
Figure \ref{fig:qt} shows the typical structure of the potential $V(x)$. 
From this, one can make many qualitative discussions. 
The first observation is that, 
near the singular boundary $x=c$, 
\begin{gather}
V(x\rightarrow c) \simeq - \frac{(5b-2T-1)/4}{c-x} - \frac{1/4}{(c-x)^2} .
\label{eq:qttc}
\end{gather}
%
The dominant term near $x=c$ is the famous $-\alpha/x^2$ potential \cite{essin2006quantum}.
It is well-known that if the range of $x$ is infinite, the
solution is unstable (not bounded from below, no ground state) unless $\alpha < 1/4$ and, 
when this is the case, all  eigenvalues are positive and the spectrum is continuous. 
In our case, $\alpha$ is exactly the limiting value 1/4. 
This exact coincidence is not well understood yet. It does not depend on $k$, $L$, $T$, 
or any specific form of the metric function $f(r)$. 
It acts as a topological constant, inherent in the geometric structure of the BH spacetime.

Because of this exact $1/4$ value, rigorous mathematical study \cite{zettl2012sturm} shows that 
our singular S--L equation can be regularized to an equivalent regular S$-$L one.
Thus, all general properties of the regular S$-$L equation are applicable in our case. 
Most importantly, the spectrum of eigenvalues, $\mu_n^2$, are discrete and countable, 
unbounded from above and bounded from below, 
as already given in Eq. (\ref{eq:munlimit}).
The corresponding eigenfunction $\omega_n(x)$ has exactly $n$ zeros in $(0,c)$. 
Therefore, $\mu_n$ are the threshole chemical potentials for the ground, 1st excited, ... states that we are seeking.
The regularization procedure requires introduction of regularization parameters
and regularized boundary conditions at $x=c$.
However, they are regularization-scheme dependent, irrelevant to our problem, 
and do not influence physical quantities. 
Similar conclusions are also obtained when investigating ``pure'' $-1/x^2$ potential \cite{essin2006quantum}.

For the second observation, 
one sees from Fig. \ref{fig:qt} that $V(x)$ is bounded from above, 
$V(x) \le V_{\mbox{\scriptsize max}}$. Therefore, one can obtain analytical expressions in the limit of asymptotically large eigenvalues $\mu_n$. 
Indeed, at $\mu^2 \gg V_{\mbox{\scriptsize max}}$, the Schr\"{o}dinger equation becomes: 
\begin{equation}
\omega_{xx} + \mu^2 \omega = 0 .    
\end{equation}
This is the Schr\"{o}dinger equation for a free particle in an infinite square-well potential. The general solution is $\omega(x) = A e^{i\mu x} + B e^{-i\mu x}$. Since the range of $x$ is finite, the wave number is quantized:
\begin{equation}
\mu_n = n \frac{\pi}{c} + \mbox{const.} 
\label{eq:mun}
\end{equation}
This is our most important result. To the leading order, this spectrum does not depend on the specific potential $V(x)$ or specific boundary conditions.
These details only contribute a small (position-dependent)
phase shift in the eigenfunctions, and is different for different types of condensates.

Remind that $c$ diverges logarithmically with $1/T$.
As $T\rightarrow 0$, Eq. (\ref{eq:mun}) shows that the spectrum of eigenvalues with high quantum number $n$ remains discrete, and equal--spacing, but the gap between $\mu_n$ shrinks to an infinitesimal value.
All these properties are in good agreement with numerical results and 
explain non-saturation behaviors observed in Fig. \ref{fig:mucvsT}. 
Surprisingly, this asymptotic behavior seems to apply even for relatively low $n$ in those figures, 
as low as $n \ge 1$. 
Note that, for the case $L=9$ (the gas side of the vdW BH transition), 
threshold values $\mu$ for three lowest energy states immediately saturate, 
the gap between them does not shrink with decreasing $T$.
On the other hand, remarkably for the case $L=1$ (the liquid side of the vdW BH transition),
the gap between the 1st excited state and the ground state keeps shrinking. 
This might correlate with the fact that we do not see gap in the conductivity 
at low temperature for excited states in the gas side of the vdW transition
\cite{phat2021triplet}.
In both cases, the ground state threshold $\mu_0$ saturates at a finite, non-zero value.

The consequence of this observation is more profound. 
We postulate that at the exact limit $T = 0$ where {\em the gap at large $n$ is zero}, all higher excited states of the condensate are activated beyond a finite chemical potential (approximately $\mu_0$ or $\mu_2$ depending on liquid or gas side of the BH transition), suggesting a potentially new quantum phase transition as function of the chemical potential. 
Unfortunately, the $T=0$ limit is non-analytical and 
more investigations are needed to confirm this.
This continuous spectrum at the $T\rightarrow 0$ limit has largely been missed in previous studies of the excited states of the holograhic condensate.

For the final observation, 
Eq. (\ref{eq:qttc}) shows that the range of the singular $-1/x^2$ term is
within $1/(5b-1)$ of the ``origin'' $x=c$. 
As $T\rightarrow 0$, $c\rightarrow\infty$, 
this localized region becomes parametrically small. 
Therefore, we regard this term as a singular potential that ``pins'' the wave function to zero at the origin, $\omega(c)=0$. 
Away from this parametrically small region, the Coulomb-like $-1/{x}$ term in Eq. (\ref{eq:qttc}) dominates.
Its spectrum is well-known \cite{yepez1987one}.
In one-dimension, the Schr\"{o}dinger equation for the hydrogen atom have discrete negative eigenvalues 
\begin{equation}
\mu_n^2 \simeq - \frac{1}{64n^2} {(5b-1-2T)^2} ~.
\end{equation}
To avoid negative values for $\mu^2$ in the ground state in the original problem,
one needs the constant shift $V_{\mbox{\scriptsize max}} \ge (5b-1)^2/64$,
which is satisfied for all $k$, $L$ parameters 
\footnote{There is an interesting condition $5b>1$ or $Lk < 2/\sqrt{5}$. 
 For $Lk > 2/\sqrt{5}$, the Coulomb term is repulsive.},
 thus there are no negative eigenvalues in our problem. 

\section{Discussion}

\label{sec:disc}

The main result of this work is the discovery of the continuous spectrum of excited states at large chemical potential of the holographic supercondutor in AdS/CFT duality at zero temperature.  
In this section, let us discuss some additional aspects of our theory.
%
Although the 4D Maxwell-Einstein black hole at finite topoligical charge is chosen to be investigated, 
our derivation is general enough to be easily expanded to other types of black holes
with different function $f(r)$ of the metric,  and in different dimension. 
Indeed, a quick check with other models of BH easily show that, 
our qualitative conclusion of a continuous spectrum of excited states at $T=0$ remains unchanged.
This fact need to be carefully taken into account
in any investigations of the excited states of holographic superconductors in AdS/CFT duality

An important step toward obtaining this general derivation is the factorization of the $f(r)$ function 
as shown in Eq. (\ref{eq:frcompact}). In the original problem, there are five independent parameters, $k$, $L$,
$M$, $Q$, and $q$. Setting $Q=q=1$, one has three independent parameters, $k$, $L$, and $M$ 
(or equivalently, $k$, $L$, $T$).
In practice, we rescale all the quantities to dimensionless, with the blackhole horizon at $z=1$.
This effectively reduces the number of parameters to only two, $b$ and $T$, which make the physical problem
much cleaner, and more tractable. 
Especially, after factorizing out the $(z-1)$ factor in $f(r)$, 
the left-over factor becomes continuous and positive in all of the $z \in [0,1)$ interval for any finite $T>0$. 
This way of factorization can be done for any types of $f(r)$ function for BH, in any dimensions. 
Therefore the continuous spectrum is robust and general, 
not limitted to the 4D Maxwell - Einstein charged blackhole that we studied.

Also due to our rescaling, $k$ and $L$ varies implicitly inside the parameters $z_h$ and $T$ which are always
positive. This means all three cases of $k = -1$, 0 or 1 depending on the horizon geometry (hyperbolic, flat or spherical) do not affect the asymptotic behavior. 
Fixing $k$ only change the value $L_c$ at which the vdW--like  small to large BH transition takes place. 
Likewise, fixing $L$ only change the value $k_c$ at which the vdW transition takes place.
Additionally, because of this, the vdW transition is a secondary consideration, 
not important to the problem of asymptotic behavior, included in the consideration
in order to investigate a more complete phase diagram of the BH model.
Nevertheless, our numerical results shown in Fig. \ref{fig:mucvsT} suggest that this vdW transition 
can strongly affect the qualitative behaviors of the few lowest excited states. 
On the liquid side the vdW-like BH transition, there seems to be no gap between 
the ground and the first excited state.
On the gas side, there are more discrete states below the continuous spectrum.

In this work, the BH system is solved without back reaction. 
Back reaction is especially required for ground state scalar where it may diverge in the $T=0$ limit. 
However, it is already known that the scalar of excited states do not diverge in the zero temperature limit
\cite{wang2020excited}.
Furthermore, inclusion of back reaction does not eliminate these excited states, only leads to some
quantitative differences \cite{arxiv.1911.04475}. 
Although those studies are for BH of zero topoligical charge, $k=0$,
we believe that the inclusion of back reaction do not
change the qualitative conclusion of our work for finite $k$. 
We plan to include the effect of back reaction in a near future work to fully address this question.

Another aspect of AdS/CFT duality that should be mentioned is that we have use the AdS written 
in global coordinates. Although a direct transformation from the local coordinates based on the Poincare patch
to the glocal coordinates is known to be not available \cite{bayona2007anti},
it would be interesting to investigate the latter to understand how our results might be affected, 
and if the structure of the effective potential would change qualitatively.
This requires a seperate investigation and will be the subject of another work.

Finally, there are some previous studies investigating different phases of the global AdS black holes 
\cite{Basu_2016,Markeviciute_2016}. In these works, addition of parameters of hairy blackhole systems
leads to the appearance of many other solutions. Interestingly, among them, the boson star solutions also 
have states with equidistant spacing. These studies, however, investigate different BH models
with more complicated metric, and more parameters. The equidistant spacing between states is an accidental coincident with our threshold $\mu_n$ spectrum.  
In this work, we focus on the excited states appearing at large chemical potential which is a different physical phenomenon.

\section{The order--parameter critical exponent}

\label{sec:criticalExp}

As one last comment, let us study the system below but close to the critical temperature to obtain the critical exponent.
Our derivation follows a similar scheme as that of Ref. \cite{siopsis2010analytic}, adapted to our boundary conditions.

Setting $\Psi_n(z) = \avg{{\cal O}_{\delta}} z^{\delta-1} \eta_n(z)$,
from the equation for $\Phi(z)$ we have
\begin{gather}
\Phi'' = {2\Psi_n^2}\Phi /F  = {2 \avg{{\cal O}_{\delta}}^2 z^{2(\delta-1)} \eta_n^2} \Phi /F .
\label{eq:phi}
\end{gather}
Near $T_c$, $\avg{{\cal O}_\delta}$ is very small, we expand in this parameter:
\begin{gather}
\Phi(z) \simeq \mu(1-z) + \avg{{\cal O}_\delta}^2 \chi(z) ,
\end{gather}
with $\chi(0)=\chi(1)=0$, $\chi(z) < 0$. Substituting into Eq. (\ref{eq:phi}), keeping only the leading terms, one gets
\begin{gather}
\chi''(z) =  \mu (1-z) {2 z^{2(\delta-1)} \eta_n^2(z)}/{F(z)} ,
\end{gather}
After integration, one has
\begin{gather}
\chi'(z) 	= \chi'(0) + \mu  \int_0^z ds ~ \frac{2 s^{2(\delta-1)} \eta_n^2(s)}{F(s)/(1-s)} ,
\end{gather}
so
\begin{gather}
\chi(z) 
 = \int_0^z dz_1~ \chi'(z_1) 
    = \chi'(0) z + \mu \int_0^z d z_1 \int_0^{z_1} ds ~ \frac{2 s^{2(\delta-1)} \eta_n^2(s)}{F(s)/(1-s)} .
\end{gather}
The condition $\chi(1)=0$ gives
\begin{gather}
\chi'(0) = - {\cal C}_n = \mu \int_0^1 d z_1 \int_0^{z_1} ds ~ \frac{2 s^{2(\delta-1)} \eta_n^2(s)}{F(s)/(s-1)} .
\end{gather}
By this definition, the constant ${\cal C}_n$ is finite and positive. From the limit $\Phi(z\rightarrow 0) \simeq \mu - \rho z$, we have
\begin{gather}
\mu - \rho z
    =  \mu(1-z) - \avg{{\cal O}_\delta}^2 {\cal C}_n z  , \\
\avg{{\cal O}_\delta} 
    \simeq \sqrt{{\mu}/{{\cal C}_n}} 
    ~ \left(1 - \frac{T}{T_c}\right)^{1/2} ~ .
\label{eq:Odelta}
\end{gather}
Here we use the fact that near $T_c$, $\rho \simeq \mu T_c/T$.
Eq. (\ref{eq:Odelta}) shows that the condensates obey universal mean-field critical exponent 1/2 for all ground and excited states.

\section{Conclusion}

\label{sec:conclusion}

In conclusion, by considering holographic phase transition in the AdS/CFT duality in the probe limit, 
excited states of the condensate of the conformal field living at the AdS boundary are investigated
in the case of asymptotically large chemical potential.
The spectrum of threshold chemical potentials for the appearance of the states is discrete, 
equal spacing, with the gap between consecutive values shrinks to zero logarithmically in the limit $T\rightarrow 0$. 
Surprisingly, numerical solution for 4D Maxwell - Einstein charged black hole at finite topological charge shows that this asymptotic behavior is valid for states as low as the first or second excited states.
This is especially significant on the liquid side of the van der Waals - like small-large black hole transition,
where there seems to be no gap between the ground and first excited state at zero temperature.

It is postulated that, at the exact limit $T=0$ where the gap vanishes, 
all excited states are activated beyond a finite chemical potential, 
suggesting a potentially new quantum phase transition.
This continuous spectrum of excite states of the condensate at the
$T\rightarrow 0$ limit has largely been missed in previous studies
of excited states of the condensate, and a more comprehensive investigation
to fully understand this new quantum phase transition.

\acknowledgments

This work is supported by the Vietnam National Foundation for Science and Technology Development under the Grant No. 103.01-2017.300.
We thank Drs. D. T. Son, Cao H. Nam, P. H. Lien, and H. V. Quyet for useful discussions.


\bibliographystyle{JHEP}      
\bibliography{asympBH}   

%
%
%
%
%
%
%
%

\end{document}